\documentclass{article}
\usepackage{bookmark}
\usepackage{microtype}
\usepackage{graphicx}
\usepackage{subfigure}
\usepackage{booktabs} 

\newtheorem{assumption}{Assumption}

\newtheorem{Lemma}{Lemma}
\providecommand{\tabularnewline}{\\}

\makeatother

\usepackage[english]{babel}

\usepackage{amsmath}

\usepackage{amsfonts}
\usepackage{mathtools}
\usepackage{xcolor,color}
\newcommand{\new}[1]{\textcolor{black}{#1}}

\newcommand{\simiid}{\stackrel{\mathrm{iid}}{\sim}}

\renewcommand{\v}[1]{\boldsymbol{#1}}

\newcommand{\bb}[1]{\mathbb{#1}}

\newcommand{\mi}{{\mathrm{i}\hspace{0.02cm}}}

\usepackage[accepted]{icml2020arxiv}

\icmltitlerunning{Spectral Subsampling MCMC for Stationary Time Series}

\begin{document}

\twocolumn[
\icmltitle{Spectral Subsampling MCMC for Stationary Time Series}




\begin{icmlauthorlist}
\icmlauthor{Robert Salomone}{UNSW}
\icmlauthor{Matias Quiroz}{UTS}
\icmlauthor{Robert Kohn}{UNSW}
\icmlauthor{Mattias Villani}{SU,LIU}
\icmlauthor{Minh-Ngoc Tran}{USYD}
\end{icmlauthorlist}

\icmlaffiliation{UNSW}{UNSW Sydney}
\icmlaffiliation{UTS}{University of Technology Sydney}
\icmlaffiliation{SU}{Stockholm University}
\icmlaffiliation{LIU}{Link\"oping University}
\icmlaffiliation{USYD}{University of Sydney}

\icmlcorrespondingauthor{Robert Salomone}{r.salomone@unsw.edu.au}

\icmlkeywords{Markov chain Monte Carlo, , Large datasets, Spectral methods}

\vskip 0.3in
]



\printAffiliationsAndNotice{}  

\begin{abstract}
    Bayesian inference using Markov Chain Monte Carlo (MCMC) on large datasets has developed rapidly in recent years. However, the underlying methods are generally limited to relatively simple settings where the data have specific forms of independence. We propose a novel technique for speeding up MCMC for time series data by efficient data subsampling in the frequency domain. For several challenging time series models,  we demonstrate a speedup of up to two orders of magnitude while incurring negligible bias compared to MCMC on the full dataset. We also propose alternative control variates for variance reduction based on data grouping and coreset constructions.
    \end{abstract}
    
    \section{Introduction}
        Bayesian inference has gained widespread use in Statistics and Machine Learning largely due to convenient and quite generally applicable Markov Chain Monte Carlo (MCMC) and Hamiltonian Monte Carlo (HMC) algorithms that simulate from the posterior distribution of the model parameters. 
        
        However, it is now increasingly common for datasets to contain millions or even billions of observations. This is particularly true for temporal data recorded by sensors at increasingly faster sampling rates. MCMC is often too slow for such big data problems and practitioners are replacing MCMC with more scalable approximate methods such as Variational Inference \citep{blei2017variational}, Approximate Bayesian Computation \citep{Marin2012} and Integrated Nested Laplace Approximation \citep{rue2009approximate}. 
        
        A recent strand of the literature instead proposes methods that speed up MCMC and HMC by data subsampling, where the costly likelihood evaluation in each MCMC iteration is replaced by an estimate from a subsample of data observations \citep{quiroz2019speeding, dang2019hamiltonian} or by a weighted coreset of data points found by optimization \citep{Campbell2018, Campbell2019, Campbell2019a}. Data subsampling methods require that the log-likelihood is a sum, where each term depends on a unique piece of data --- a condition satisfied for independent observations or for independent subjects in longitudinal data (with potentially dependent data within each subject) --- but does not hold for general time series problems.
        
        Our paper extends the applicability of previously proposed subsampling methods to stationary time series. The method is based on using the Fast Fourier Transform (FFT) to evaluate the likelihood function in the frequency domain for the periodogram data. The advantage of working in the frequency domain is that under quite general conditions the periodogram observations are known to be asymptotically independent and exponentially distributed with scale equal to the spectral density. The logarithm of this so called \emph{Whittle likelihood} approximation of the likelihood is therefore a sum even when the data are dependent in the time domain. The asymptotic nature of the Whittle likelihood makes it especially suitable here since subsampling tends to be used for large-scale problems where the Whittle likelihood is expected to be accurate. Moreover, our algorithm can also be used with the recently proposed \new{Debiased Whittle Likelihood} \citep{Sykulski2019} which gives better likelihood approximations for smaller datasets. 
        
        It is by now well established that efficient subsampling MCMC methods require likelihood estimators with low variance \citep{quiroz2019speeding, quiroz2018subsampling}. Variance reduction is typically achieved by using control variates that approximate the individual log-likelihood terms, often by assuming that these terms are approximately quadratic around a reference value. Our second contribution proposes a grouping strategy that makes the grouped log-likelihood terms more quadratic compared to that of the individual log-likelihood terms. In addition, when the number of observation in each group is large, we propose to use the coreset construction of \citet{Campbell2018} to approximate the grouped log-likelihood. This is advantageous when the grouped log-likelihood terms are not approximately quadratic.
       
        The structure of our paper is as follows. Section \ref{sec:Whittle} introduces the necessary frequency domain concepts and defines the Whittle likelihood. Section \ref{sec:SubsamplingMCMC} gives an overview of the Subsampling MCMC approach of \citet{quiroz2019speeding}. Section \ref{sec:EnhancedCV} introduces our novel control variate schemes. Section \ref{sec:Numerics} summarizes the results of experiments on examples of models that have previously not been feasible with large data methods, such as long memory stochastic volatility models.

    \section{Data Subsampling Using the Whittle Likelihood}\label{sec:Whittle}
    \subsection{Discrete Fourier Transformed Data}
    Let $\{X_t\}_{t=1}^n$ be a covariance stationary zero-mean time series with covariance function $\gamma(\tau) \coloneqq \mathbb{E}X_t X_{t-\tau}$ for $\tau \in \mathbb{Z}$. The {\em spectral density} is the Fourier transform of $\gamma(\tau)$ \citep{lindgren2012stationary}
    \begin{equation}\label{eq:spectralDens}
        f(\omega) = \frac{1}{2\pi}\sum_{\tau = -\infty}^\infty \gamma(\tau)\exp(-\mi \omega\tau),
    \end{equation}
    where $\omega \in (-\pi,\pi]$ is called the {\em angular frequency}. The {\em discrete Fourier Transform} (DFT) of $\{X_t\}_{t=1}^n$ is the complex valued series 
    \begin{equation}\label{eq:DFT}
        J(\omega_k) \coloneqq\frac{1}{\sqrt{2\pi}} \sum_{t=1}^n X_t \exp(-\mi \omega_k t),
    \end{equation}
    for $\omega_1,\ldots,\omega_n$ in the set of Fourier frequencies
    $$\Omega = \{2\pi k/n, \text{ for } k = -\lceil n/2\rceil + 1,\ldots,\lfloor n/2 \rfloor \}.$$
    The DFT is efficiently computed by the Fast Fourier Transform (FFT). The {\em periodogram} ${\cal I}(\omega_k) \coloneqq n^{-1}\vert J(\omega_k) \vert ^2$ is an asymptotically unbiased estimate of $f(\omega_k)$.
    
    \subsection{Frequency Domain Asymptotics}\label{subsec:FreqDomainAsymptotics}
    The DFT in \eqref{eq:DFT} is a linear transformation that acts like a weighted average of time-domain data. A central limit theorem can therefore be used to prove that $J(\omega_k)$ are asymptotically \new{independent} complex Gaussian under quite general conditions \citep[Corollary 2.1]{Shao2007}; see also \citet[Theorem 2.1]{peligrad2010central} who establish the result for $\omega \in (0, 2\pi)$ almost-everywhere under even weaker conditions.
    
    Furthermore, the real and imaginary parts are asymptotically independent. Denote a chi-squared random variable with $r$ degrees of freedom as $\chi_r^2$. The scaled periodogram ordinate ${\cal I}(\omega_k)/f(\omega_k)$ is asymptotically distributed as $\chi^2_2 /2$ (i.e., standard exponential) for all $k\neq0,n$; and as $\chi^2_1$ for $k=0$ and $k=n$. Hence, we have the following asymptotic distribution of the periodogram
    \begin{equation}\label{eq:asympPeriodogram}
        {\cal I}(\omega_k)\ {\sim}\ {\sf Exp}(f(\omega_k)), \quad k=1,\ldots,n-1 
    \end{equation}
     independently as $n\rightarrow \infty$, with the exponential distribution in the scale parameterization, i.e., parameterized by its mean. 
  
    \subsection{The Whittle Likelihood}
    The asymptotic distribution of the periodogram ordinates in \eqref{eq:asympPeriodogram} motivates the Whittle log-likelihood \citep{Whittle1953} for a time series model with parameter vector $\v\theta$:
    \begin{equation}\label{eq:theWhittleLikelihood}
        \ell_W({\v \theta}) = -\sum_{k=1}^{\lfloor (n-1)/2 \rfloor} \Big( \log f_{{\v \theta}}(\omega_k) + \frac{{\cal I}(\omega_k)}{f_{{\v \theta}}(\omega_k)} \Big),
    \end{equation}
    where $f_{\v \theta}(\omega)$ is the spectral density of the model. For real-valued data, both $f_{{\v \theta}}(\omega_k)$ and ${\cal I}(\omega_k)$ are symmetric about the origin, so the Whittle log-likelihood is evaluated by summing only over the non-negative frequencies; for demeaned data, the term for $\omega_k=0$ is removed from the likelihood as then $J(\omega_k)=0$.
    
    The Whittle log-likelihood has several desirable properties that enable scalable Bayesian inference:
    
    \begin{itemize}
        \item The periodogram does not depend on the parameter vector $\v \theta$ and can therefore be computed before the MCMC at a cost of ${\cal O}(n \,{\rm log}\, n)$ via the Fast Fourier Transform algorithm. After this one-time cost, likelihood evaluations have the same ${\cal O}(n)$ cost as for independent data. 
        \item The Whittle log-likelihood is a sum in the frequency domain and is therefore amenable to subsampling using the {\em same} algorithms developed for independent data in the time domain.
        \item As the Whittle log-likelihood relies on large sample properties of the periodogram, it is particularly suited to large datasets where subsampling MCMC and related methods are used.
    \end{itemize}
    
    Below, the term log-likelihood refers to the Whittle log-likelihood, and $n$ to be the number of unique summands in \eqref{eq:theWhittleLikelihood} --- i.e., we assume the FFT has been performed and we are working in the frequency domain. 
    
    \section{Subsampling MCMC}\label{sec:SubsamplingMCMC}
   The fundamental idea in the previous section can be used to extend {\em any} existing method for subsampling that requires conditionally independent data to the case of fitting a parametric stationary time series model with known spectral density. To provide a proof-of-concept in the form of numerical experiments, we focus on the Subsampling MCMC approach of \citet{quiroz2019speeding}, as it has been shown to give more accurate posterior inferences than other approaches such as Stochastic Gradient Langevin Dynamics (SGLD) \citep{SGLD} and Stochastic Gradient Hamiltonian Monte Carlo (SG-HMC) \citep{SG-HMC}), see for example \citet{dang2019hamiltonian}.

   We also propose novel control variates that we use with Subsampling MCMC which are also likely to be useful in further improving SG-HMC and SGLD.  
 
    \subsection{MCMC with an estimated likelihood}
    
    Let $\pi({\v \theta}) \propto L_n({\v \theta})p({\v \theta})$ denote the posterior distribution from a sample of $n$ observations with likelihood function $L_n({\v \theta})$. MCMC and HMC algorithms sample iteratively from $\pi({\v \theta})$ by proposing a parameter vector ${\v \theta}^{(j)}$ at the $j$th iteration and accepting it with probability
    \begin{equation}\label{eq:MHaccProb}
        \min\Bigg\{1,
        \frac{L_n(\v\theta^{(j)})p(\v\theta^{(j)})}{L_n(\v\theta^{(j-1)})p(\v\theta^{(j-1)})}
        \cdot
        \frac{g(\v\theta^{(j-1)} \vert \v\theta^{(j)})}{g(\v\theta^{(j)} \vert \v\theta^{(j-1)})}
        \Bigg\},
    \end{equation}
    where $g(\cdot\vert\cdot)$ is the proposal distribution.
    Repeated evaluations of the likelihood in the acceptance probability are costly when $n$ is large. \citet{quiroz2019speeding} propose speeding up MCMC for large $n$ by replacing $L_n(\v \theta)$ with an estimate $\widehat L(\v \theta,\mathbf{u})$ based on a small random subsample of $m\ll n$ observations, where $\mathbf{u}=(u_1,...,u_m)$ indexes the selected observations. 
    
    Their algorithm samples $\v\theta$ and $\mathbf{u}$ jointly from an extended target distribution $\tilde \pi (\v\theta,\mathbf{u})$. \citet{Andrieu2009} prove that such \emph{pseudo-marginal MCMC} algorithms sample from the full-data posterior $\pi({\v \theta})$ if the likelihood estimator is unbiased; i.e., $\mathbb{E}_{\mathbf{u}}\widehat L(\v \theta,\mathbf{u}) =  L({\v \theta})$.
    
    \citet{quiroz2019speeding} use an unbiased estimator of the log-likelihood $\widehat \ell(\v\theta,\mathbf{u})$ and subsequently debias $\exp(\widehat \ell(\v\theta,\mathbf{u}))$ to estimate the full-data likelihood. Although the debiasing approach in general cannot remove all bias, their pseudo-marginal sampler is still a valid MCMC algorithm, targeting a slightly perturbed posterior which is shown to be within $O(n^{-1}m^{-2})$ distance in total variation norm of the true posterior. See \citet{Quiroz2016} for an alternative {\em completely} unbiased likelihood estimator and \citet{dang2019hamiltonian} for an HMC extension. 
    
    \subsection{Estimators based on control variates}\label{subsec:ControlVariates}
    Assume that the log-likelihood decomposes as a sum $\ell(\v \theta) = \sum_{k=1}^n \ell_k(\v\theta)$; either by assuming independent data or by using the Whittle likelihood in the frequency domain for temporally dependent data. A naive estimator of the log-likelihood is  
     \[
         \widehat{\ell}_{\rm naive}(\v \theta) \coloneqq \frac{n}{m}{\sum_{i=1}^m\ell_{u_i}}(\v \theta),
     \]
     where $u_1,\ldots,u_m \simiid {\sf Unif}(\{1,\ldots, n\})$ and we suppress dependence on $\v u$ in the notation for $\widehat{\ell}$ for notational clarity. This estimator typically has large variance and is prone to occasional gross overestimates of the likelihood causing the MCMC sampler to become stuck for extended periods and thus become very inefficient.
     
     \citet{quiroz2019speeding} propose using a control variate to reduce the variance in the so-called {\em difference estimator}
      \begin{equation}\label{eq:diffestimator}
          \widehat{{\ell}}_{{\rm diff}}(\v \theta) \coloneqq \sum_{k=1}^n q_k (\v \theta)
        + \frac{n}{m}\sum_{i=1}^m\bigg({\ell}_{u_i}(\v \theta) - q_{u_i}\left(\v \theta\right)\bigg),
      \end{equation}
      where $u_1,\ldots, u_m \simiid {\sf{Unif}(\{1,\ldots,n\})}$. The $q_{k}(\v\theta)$ is the control variate for the $k$th observation. It is evident from \eqref{eq:diffestimator} that the variance of $\widehat{{\ell}}_{{\rm diff}}$ is small when the $q_{k}(\v\theta)$ approximates $\ell_k(\v\theta)$ well.
      \citet{quiroz2019speeding} follow \citet{bardenet2017markov} and use a second order Taylor expansion of $\ell_k(\v\theta)$ around some central value $\v\theta^\star$ as the control variate:
      \[\begin{split}
      q_k(\v \theta) \coloneqq {\ell_k}(\v \theta^\star) + &
      \nabla_{\v \theta}{\ell_k}(\v \theta^\star)^\top(\v \theta - \v \theta^\star) 
      \\ & + \frac{1}{2}(\v \theta - \v \theta^\star)^\top \, \nabla^2_{\v \theta}{\ell_k}(\v \theta^\star)(\v \theta - \v \theta^\star).\end{split}\] 
    One advantage of this control variate is that the otherwise $O(n)$ term $\sum_{i=1}^n q_i (\v \theta)$ can be computed at $O(1)$ cost; see \citet{bardenet2017markov}.
    
    \subsection{Block Pseudo-Marginal Sampling\label{subsec:BlockPM}}
    The acceptance probability in \eqref{eq:MHaccProb} reveals that it is actually the variability of the \emph{ratio} of estimates at the proposed and current draw that matters for MCMC efficiency \citep{deligiannidis2018correlated}. \citet{Tran2016block} propose a blocked pseudo-marginal scheme for subsampling that partitions the indicators in $B$ blocks $\mathbf{u} = (\mathbf{u}_1,...,\mathbf{u}_B)$ and only updates one of the blocks in each MCMC iteration. Under simplifying assumptions, Lemma 2 in \citet{Tran2016block} shows that blocking induces a controllable correlation between subsequent estimates in the MCMC of the simple form
    $$\mathrm{Corr}\big(\widehat \ell(\v\theta^{(j)}),\widehat \ell(\v\theta^{(j-1)})\big) \approx 1-1/B.$$
  
    \section{Alternative Control Variates via Grouping}\label{sec:EnhancedCV}
 
    The control variates presented in Section \ref{subsec:ControlVariates} are only good approximations if the {\em individual} log-likelihood terms, ${\ell_k(\v \theta)}$, are approximately quadratic or $||\v \theta - \v \theta^\star||$ is sufficiently small. We propose two new control variates that may be preferable when this is not the case. 
    
    \subsection{Grouped Quadratic Control Variates}\label{subsec:GQCV}
    Rather than sampling individual observations, we can sample observations in {\em groups}. The advantage of sampling groups is that the quadratic control variates are expected to be more accurate for the group as a whole compared to individual observations. The reason is that the Bernstein-von Mises theorem (asymptotic normality of the posterior) suggests an approximately quadratic log-likelihood for the group provided the number of observations in the group is large enough; see \citet{Tamaki2008} for a Bernstein-von Mises theorem specifically for the Whittle likelihood.
    
    Let $\cal G$ be a partition of the set of indices ${\cal U}=\{1,...,n\}$ into $|{\cal G}|$ groups $G_1,\ldots, G_{|{\cal G}|}$; i.e. ${\cal U} = \cup_{k=1}^{|{\cal G}|}G_k$, where $G_k$ is the set of data indices associated with the $k$th group. Similarly, write
    \[{\ell}_{G_k}(\v \theta) \coloneqq \sum_{i\in G_k}{\ell}_i(\v \theta)\]
    for the sum of log-likelihood terms corresponding to the observations in the $k$th group, noting that  ${\ell} = {\ell}_{\cup_k G_k} = \sum_k {\ell}_{G_k}$. Since $\ell_{G_k}(\v \theta)$ is based on $|G_k|$ observations we expect it to be closer to a quadratic function than the $\ell_i(\v \theta)$ belonging to the individual samples in the group. Now, define the control variate for group ${G_k}$ as
    \[\begin{split}
      q_{G_k}(\v \theta) \coloneqq {\ell_{G_k}}(\v \theta^\star) + &
      \nabla_{\v \theta}{\ell_{G_k}}(\v \theta^\star)^\top(\v \theta - \v \theta^\star) 
      \\ & + \frac{1}{2}(\v \theta - \v \theta^\star)^\top \, \nabla^2_{\v \theta}{\ell_{G_k}}(\v \theta^\star)(\v \theta - \v \theta^\star),\end{split}\]
      where the same $\theta^\star$ is used for all groups.
    The {\em grouped difference estimator} for a sample of $m$ groups is then
    \begin{equation}\label{eq:groupedDiffEst}
    \widehat{\ell}_{{\rm gr}}(\v \theta) \coloneqq \, \sum_{k=1}^{|{\cal G}|}q_{G_k}(\v \theta) + \frac{{\cal |G|}}{m}\sum_{i=1}^m\bigg({\ell}_{G_{u_i}}(\v \theta) - q_{G_{u_i}}(\v \theta)\bigg),
    \end{equation}
    where $u_1,\ldots, u_m \simiid {\sf{Unif}}(\{1,\ldots,|{\cal G}|\})$.
    
     \subsection{Grouped Coreset Control Variates}
     
    When the grouped log-likelihoods are far from quadratic, we propose an alternative method using \emph{Bayesian Coresets} \citep{Huggins2016} to construct control variates. An advantage of this approach is that it is unnecessary to select a central point $\v \theta^\star$, or to rely on a quadratic expansion function which may be unsuitable.
    
    Bayesian Coresets replace the true log-likelihood ${\ell}(\v \theta)$  with the approximation
    $\ell_{C}(\v \theta) = \sum_{k=1}^n w_k {\ell}_k(\v \theta)$,
    where $\v w$ is a {\em sparse} vector with a number of non-zero elements that is much less than $n$. Let  $\widehat{\pi}$ denote some {\em weighting distribution} that has the same support as the posterior and can easily be sampled from --- for example a Gaussian based on Laplace approximation, or the empirical distribution of samples from an MCMC on a smaller data set.
    The log-likelihood approximation $\ell_{\rm C}$ is constructed using a greedy algorithm, which, after $M$ steps, provides an approximate solution to
     \[ {\rm argmin}_{\v w \in \bb R^{n}} \left\{\bb E_{\widehat{\pi}}\left[\left({\ell(\v \theta)} - {\ell}_{C}(\v \theta)\right)^2  \right]\right\}\]
     subject to the constraints that $w_i \ge 0$ for $i=1,\ldots, n$, and $\sum_{k=1}^n \bb I\{w_k > 0\} \le M$, where $M$ is a user-specified number of iterations in the coreset optimization procedure. We use the {\em Greedy Iterative Geodesic Ascent} (GIGA) method in \citet{Campbell2018} for tackling the optimization.
    
    We propose approximating the group log-likelihoods $\ell_{G_k}(\v \theta),k=1,\ldots,\cal |G|$, by a separate coreset approximation for each group. We can use the grouped difference estimator in \eqref{eq:groupedDiffEst} with coreset approximations as control variates for each respective group. The coreset control variates are attractive as {\em by design} they approximate $\ell_{G_k}(\v \theta)$ well for each group using less density evaluations than the number of observations in the group. While the construction of coreset control variates requires $|{\cal G}|$ runs of the coreset procedure, each is only on a group of the dataset, so the overall effort is roughly that of the standard coreset approach, or less if run in parallel.
    
    \subsection{Perturbation of Subsampled Whittle Posterior}
In this section, we present a result regarding the perturbation of the subsampled Whittle posterior to the exact Whittle posterior.
  
  Let $\pi_{n}(\v \theta)\propto L_n(\v \theta)p(\v \theta)$ be the posterior based on the Whittle likelihood $L_n(\v \theta)=\exp(\ell_n(\v \theta))$ with $n$ samples. Following \citet{quiroz2019speeding} we define $\overline{\pi}_{n,m}(\v \theta,\v u)$ as the target for $\v \theta$ extended with the $m$ (group) subsample indicators $\v u$, and use MCMC to sample $\v \theta$ and $\v u$ jointly from the extended target. The MCMC algorithm produces valid draws from the marginal $\overline{\pi}_{n,m}(\v \theta)=\int\overline{\pi}_{n,m}(\v \theta,\v u)d\v u$. Note that $\overline{\pi}_{n,m}(\v \theta)\neq {\pi}_n(\v \theta)$ as with the methodology in \citet{quiroz2019speeding}, it is not possible to eliminate all bias in $\exp(\ell_n(\v \theta))$. 
    
    However, as a direct consequence of \citet[Theorem 1]{quiroz2019speeding}, we have the following lemma, which shows that the perturbation error decreases rapidly.

   \begin{Lemma} Suppose that the regularity conditions discussed in the supplement are satisfied, and that the control variates in Section \ref{subsec:GQCV} are used, with the number of groups $m$ depending on $n$ in a manner such that  $m(n) \rightarrow \infty$ as $n \rightarrow \infty$. Then,
    \[
    \int_{\v \theta} \left|\overline{\pi}_{n,m(n)}(\v \theta)-\pi_n(\v \theta)\right|d\v \theta = O\left(\frac{1}{{n m(n)^{2}}}\right).
    \]
  Moreover, for any scalar-valued function $h$ that satisfies $\lim \sup_{n \rightarrow \infty} \bb E_{\pi_{n}}[h^2(\v \theta)]<\infty$  \[|\bb E_{\overline{\pi}_{n,m(n)}}h(\v \theta) - \bb E_{\pi_n}h(\v \theta)| = O\left(\frac{1}{{n m(n)^{2}}}\right). \]
   \end{Lemma}
   
We highlight that the required regularity conditions are essentially those of \citet{quiroz2019speeding} but where the posterior density has \eqref{eq:theWhittleLikelihood} as a (log)-likelihood function. Thus, these conditions are justified by results such as the asymptotic normality of the maximum Whittle-likelihood estimator \cite{FoxTaqqu1986} and the Bernstein-von Mises theorem for the Whittle measure \cite{Tamaki2008}.
    \section{Experiments}\label{sec:Numerics}
  
\subsection{Settings and Performance Measures}
    
    There are many ways to partition the dataset into groups for the control variates. We use the same number of samples for each group. The $k$th group is chosen by starting with the $k$th lowest frequency and then systematically sampling every $|\cal G|$ frequency after that. This way we ensure that each group contains periodogram ordinates across the entire frequency range. The homogeneity of the groups makes it possible to use the same $\v \theta^\star$ for all groups.
  
    We use the Laplace approximation of the posterior as weighting function in the coreset approximation, truncated to the region of admissible parameters. Each coreset is fitted using the GIGA algorithm for $M=200$ iterations, using $500$ random projections; see \citet{Campbell2018} for details. We note that the mode used in the Laplace approximation comes with no extra cost compared to full-data MCMC as the latter uses the mode as a starting value for the sampler and to build the covariance matrix of the random walk Metropolis proposal. Likewise, the Taylor control variates are constructed using the mode as $\v \theta^\star$. Therefore, this part of the start-up cost is assumed to be the same for all algorithms. However, both control variates have additional start-up costs compared to full-data MCMC. The coreset control variate needs to perform the GIGA optimization, which makes $M$ sweeps of the full dataset, using $Mn$ density evaluations. As discussed above, this can in practice be done in parallel for each group, where each group uses $M|G_k|$ observations. Recall that $n = \sum_k |G_k|$, which explains the cost of $Mn$ density evaluations. The Taylor control variate requires summing all the $q_k$ once (first term in \eqref{eq:diffestimator}), hence adding $n$ to the total cost. For simplicity, assume all groups have the same number of observations $|G|=|G_k|$. During run time, full-data MCMC requires $n$ density evaluations in each iteration, whereas the Taylor control variate uses $m|G|$ and the coreset control variate $|m|G + \sum_{k=1}^{|\cal{G}|}g_k$, where the second term is the cost of evaluating the summation of all $q_{G_k}(\v \theta)$, which is much faster than full-data MCMC if the coreset size $g_k$ is small in relation to $|G_k|$.
    
    We follow \citet{quiroz2019speeding} and use the {\em computational time} (CT) as our measure of performance. This measure balances the cost (number of density evaluations as discussed above) and the efficiency of the Markov chain. It is defined as
    \[{\rm CT := IF}\times \text{number of density evaluations}, \]
    where the inefficiency factor (IF) is proportional to the asymptotic variance when estimating a \new{univariate} posterior mean based on MCMC output. The IF is interpreted as the number of (correlated) samples needed to obtain the equivalent of a single independent sample. It is convenient to measure the cost using density evaluations since it makes the comparisons implementation independent. We use the {\sf CODA} package \citep{plummer2006coda} in {\sf R} to estimate IF. Our measure of interest is the {\em relative} CT (RCT) which we define as the ratio between the CT of full-data MCMC and that of the subsampling algorithm of interest. Hence, values larger than one mean that the subsampling algorithm is more efficient when balancing computing cost (density evaluations) and statistical efficiency (variance of the posterior mean estimator).
   
\subsection{Experiments}\label{subsec:Experiments}
We consider several time series models for large data in our experiments, including the recently proposed class of autoregressive tempered fractionally integrated moving average (ARTFIMA) models and several of its widely-used special cases such as ARMA and ARFIMA --- though we highlight that any model class for which the spectral density is known can be used. We also consider a stochastic volatility model with an underlying ARTFIMA process. 

\citet{Sabzikar2019} defines $Y_t$ as an ${\rm ARTFIMA}(q,d,\lambda,p)$ process if
    \[\phi_{q}(L)\Delta^{d,\lambda}(Y_{t}-\mu)=\theta_p(L)\varepsilon_{t}, \]
    where $\{\varepsilon_t\}_{t \in \bb Z}$ is an iid sequence of zero mean random variables with variance $\sigma^2$, $\phi_{q}(L)\coloneqq1-\phi_{1}L-\cdots-\phi_{q}L^{q}$, and $\theta_{p}(L)\coloneqq1+\theta_{1}L+\cdots+\theta_{p}L^{p}$, are the autoregressive and moving average lag polynomials, where $L$ is the lag operator, i.e. $L^k(X_t) = X_{t-k}$. The {\em tempered fractional differencing operator} is defined by
    \begin{equation*}
    \begin{split}
     \Delta^{d,\lambda}Y_t &\coloneqq (1-e^{-\lambda}L)^d Y_t  \\ &=\sum_{j=0}^\infty (-1)^j \frac{\Gamma(1+d)}{\Gamma(1+d-j)j!}e^{-\lambda j}Y_{t-j},     
    \end{split}
    \end{equation*}
     where $d$ is called the {\em fractional integration parameter} and $\lambda$ is called the {\em tempering parameter}. 
    To explain the role of the parameters $d$ and $\lambda$, note that for $\lambda=0$ and $d$ a non-negative integer, $\Delta^{d,\lambda}Y_t$ reduces to simple differencing of order $d$ and we obtain the AutoRegressive Integrated Moving Average (ARIMA) processes. Autoregressive Fractionally Integrated Moving Average (ARFIMA) \cite{Granger1980} extends this class by allowing fractional differences, i.e. $d$ need not be an integer. For $-0.5 < d < 0.5$, ARFIMA is stationary, and is of particular interest as it has long-range or long-memory dependence with an autocovariance function that dies off so slowly it is not absolutely summable,  $\sum_{k=-\infty}^\infty |\gamma(k)| = \infty$. The tempering parameter $\lambda>0$ in ARTFIMA allows for semi-long range dependence, i.e. ARFIMA-like long-range dependence for a number of lags beyond which the autocovariance decays exponentially fast.   
    
    Provided that $\lambda > 0$, $d \notin \bb Z$, and the roots of $\phi_{q}(z)$ lie outside of the unit circle in the complex plane, the ARTFIMA process is stationary \citep[Theorem 2.2]{Sabzikar2019} with spectral density 
    \begin{equation}\label{eq:ARFIMAspec}
    f(\omega)=\frac{\sigma^{2}}{2\pi}\left|1-e^{-(\lambda + i\omega)}\right|^{-2d}\left|\frac{\theta_p(e^{-i\omega})}{\phi_{q}(e^{-\lambda i\omega})}\right|^{2}.
    \end{equation}

    We follow \citet{Barndorff-Nielsen1973} and reparameterize the autoregressive parameters, $\phi_k$ for $k=1,\ldots,q$, in terms of the {\em partial autocorrelations} $\widetilde{\v \phi}_q=(\widetilde \phi_1,\ldots,\widetilde \phi_q)$. Stationarity can now be enforced by the conditions $|\widetilde{\phi_k}|<1$, for $k=1,\ldots,q$. We perform the same reparameterization to $\v \theta_q$ to obtain $\widetilde{\v \theta}_q$, which ensures the  underlying process is invertible provided that the constraint $|\widetilde{\theta}_k| < 1$ for $k=1,\ldots,p$ is satisfied. We use the priors $\tilde{\v \phi}_{q} \sim {\sf Unif}\big((-1,1)^q\big)$ and $\tilde{\v \theta}_{p} \sim {\sf Unif}\big((-1,1)^p\big)$
    
    A log-transformation is used for both $\sigma^2$ and $\lambda$, with both priors $\log(\sigma^2), \log(\lambda) \sim {\cal N}(0,1)$. For ARFIMA models, the fractional integration parameter $d$ is parametrized by a scaled Fisher transformation $\widetilde{d} \coloneqq  {\rm arctanh}(2d)$ with prior $\widetilde{d}\sim \mathcal{N}(0,1)$, which is a weakly informative on $d=0.5\tanh(\widetilde{d})$ in the region $\left(-\frac{1}{2}, \frac{1}{2}\right)$. For ARTFIMA models ($d$ not restricted to $(-0.5, 0.5)$) we set $d \sim \mathcal{N}(0,1)$.

{\bf Example 1: Vancouver Temperatures (ARMA)}

    The first example considers the ARMA model for hourly temperature data for the city of Vancouver during the years 2012 to 2017 sourced from \url{openweathermap.org}. Using the relatively simple ARMA model makes it possible to compare with the posterior obtained by MCMC using the exact time domain likelihood from the Kalman filter. We use the {\sf stl} function in {\sf R} to remove the trend and yearly seasonal from the original series. We then confirm that the series passes the Augmented Dickey-Fuller and Phillips-Perron unit-root tests for stationarity. The final time series is of length $n=44001$, yielding a likelihood with $n=2.2 \times 10^4$ frequency terms. The {\sf auto.arima} function in the {\sf Forecast} \citep{ForecastPackage} package in {\sf R} was used for model selection, yielding an ${\rm ARMA}(2,3)$ model. 
    
{\bf Example 2: Stockholm Temperatures (ARTFIMA)}

The second example fits an ARTFIMA(2,2) model to hourly temperature data for the city of Stockholm during the years 1967 to 2018, obtained from the Swedish Meteorological and Hydrological Institute (\url{www.smhi.se/en}). The data preparation procedure is the same as in the last example. The series is of length $4.5 \times 10^5 + 1$ yielding $2.25 \times 10^5$ terms in the Whittle likelihood.
  
{\bf Example 3: Simulated ARMA Series (ARFIMA)}

To assess the performance of our methods on even larger datasets, we simulate an ARMA(2,1) series of length $n=5 \times 10^6 + 1$, and fit an ARFIMA(2,1) model. The parameters of the data generating process are $\v \phi = (0.22, -0.1)$, $\v \theta = (0.5)$, and $\sigma^2 = 1$. 
    
{\bf Example 4: Bitcoin Prices Stochastic Volatility (ARTFIMA-SV)}

Finally, we test our methods on a challenging ARTFIMA model extended with stochastic volatility. A general class of Stochastic Volatility (SV) models is
\begin{equation}\label{eq:SV}
    y_t = \exp(v_t/2)\xi_t,
\end{equation}
where $\{\xi_t\}$ is an independent and identically distributed sequence having mean zero and unit variance, and $v_t$ is a stationary process with parameter vector $\v \psi$ and spectral density $f_{v}(\omega ; \v \psi)$. \citet{BREIDT1998325} observe that the SV model in \eqref{eq:SV} can be estimated by noting that 
\begin{equation}\label{eq:SVtransform}
\log y_t^2 = \mu + v_t + \varepsilon_t,    
\end{equation}
where $\mu \in \bb R$, and $\{\varepsilon_t\}$ is independent and identically distributed white noise process with zero mean and variance $\sigma_{\varepsilon}^2$. 
Thus, as the spectral density of $\varepsilon_t$ is $f_{\varepsilon}(\omega) = \sigma_{\varepsilon}^2/2\pi$, 
fitting a parametric spectral density of the form
\[f(\omega ; \v \psi, \sigma_{\epsilon}^2) = f_{v}(\omega ; \v \psi) + \frac{\sigma_{\epsilon}^2}{2 \pi},\]
to the log-squared series is equivalent to fitting the model in \eqref{eq:SV} to the original series. We set $\log \sigma_{\varepsilon}^2 \sim \mathcal{N}(0,0.01)$ a priori.

We let $v_t$ be an ARTFIMA(1,1) process, which generalizes the Long Memory Stochastic Volatility model of \citet{BREIDT1998325} to allow for tempered fractional differencing. This is a challenging model to estimate even without tempering since the computational cost of filtering to obtain the Gaussian ARFIMA likelihood scales poorly with the length of the time series \cite{chan1998state}. Tempering introduces additional difficulty as the ARTFIMA covariance function involves infinite sums involving the Gaussian hypergeometric function \citep[Equation (2.9)]{Sabzikar2019}. We show that spectral subsampling MCMC is a computationally efficient way to obtain the posterior of the ARTFIMA-SV model for large datasets. We fit the model to a dataset of one-minute Bitcoin returns (prices from the exchange on {\url{coinbase.com}}) of length $n=10^6 + 1$ (resulting in $n=5\times 10^5$ terms in the Whittle likelihood).

\begin{table}[ht]
\footnotesize 
      \centering
\begin{tabular}{lcrrrrrrrc}
\hline
 &  & $\mathcal{M}_1$ &  & $\mathcal{M}_2$ &  & $\mathcal{M}_3$ &  & $\mathcal{M}_4$ & \tabularnewline
\cline{3-3} \cline{5-5} \cline{7-7} \cline{9-9} 
$n~(\times10^{4})$  &  & $2.2$ &  & $22.5$ &  & $250$ &  & $50$ & \tabularnewline
$|\mathcal{G}|~(\times10^{3})$ &  & $1$ &  & $1$ &  & $1$ &  & $1$ & \tabularnewline
$|G|$ &  & $22$ &  & $225$ &  & $2500$ &  & $500$ & \tabularnewline
$m$ (\%$|\mathcal{G}|$) &  & $2$ &  & $1$ &  & $1$ &  & $1$ & \tabularnewline
$B$ &  & $10$ &  & $10$ &  & $10$ &  & $10$ & \tabularnewline
$M$ &  & - &  & $200$ &  & $200$ &  & $200$ & \tabularnewline
$RP$ &  & - &  & $500$ &  & $500$ &  & $500$ & \tabularnewline
$\bar{g}_{k}$ &  & - &  & $29$ &  & $34$ &  & $39$ & \tabularnewline
\hline 
\end{tabular}
      \caption{Settings for models ARMA(2,3) ($\mathcal{M}_1$), ARTFIMA(2,2) ($\mathcal{M}_2$), ARFIMA(2,1) ($\mathcal{M}_3$) and ARTFIMA-SV(1,1) ($\mathcal{M}_4$). The table shows number of frequency observations ($n$), number of groups ($|\mathcal{G}|$), number of observations per group ($|G|$, the same for all groups), percentage of subsampled groups ($m$) and number of blocks in the block pseudo-marginal algorithm ($B$). The coreset settings are number of iterations of GIGA algorithm ($M$), number of random projections ($RP$) (see \citet{Campbell2018} for details) and the average size of the coreset ($\bar{g}_k$).}
      \label{tab:settings}
  \end{table}
  
      \begin{figure}[ht]
        \centering
        \includegraphics[width=0.50\textwidth]{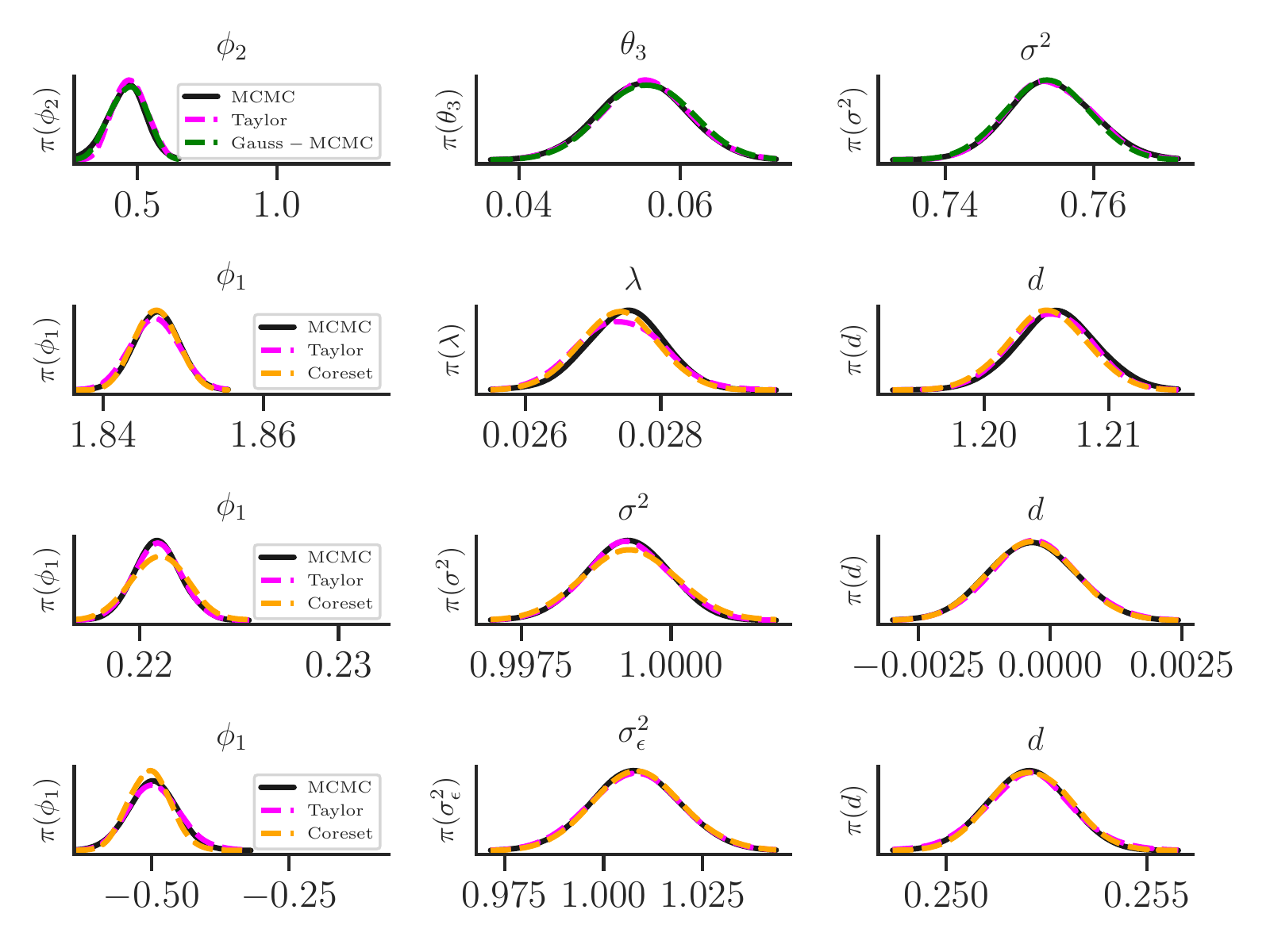}
        \caption{Kernel Density Estimates of some marginal distributions (see the supplementary for all parameters) for all examples. Each row corresponds to a single model (from the top: ARMA(2,3), ARTFIMA(2,2), ARFIMA(2,1) and ARTFIMA-SV(1,1)). MCMC is the full-data MCMC on the Whittle likelihood. Taylor and coreset are the Subsampling MCMC methods using the corresponding control variates. Gauss-MCMC is the full-data MCMC with the time-domain Gaussian likelihood.  } \label{fig:kde_all_models}
    \end{figure}

\subsection{Results}
 Table \ref{tab:settings} displays the settings for each example. Note that only for the ARMA(2,3) model is it computationally feasible to compare with the posterior based on the exact time domain likelihood. Furthermore, the number of observations per group in the Vancouver temperature data is too small (22) for the coreset control variates to be useful. 
 
 Figure \ref{fig:kde_all_models} displays a selection of kernel density estimates for marginal posteriors across all examples. Note that the incurred bias from subsampling is neglible, especially for the Taylor series control variate. The coreset control variate results in a higher variance and thus a larger perturbation error \cite{quiroz2019speeding}.  Figure \ref{fig:kde_all_models} also shows that the posterior based on the Whittle likelihood is close to the Gaussian time-domain posterior for the ARMA(2,3) example. We stress that we can not make this comparison for the other models because, as discussed in Section \ref{subsec:Experiments}, the Gaussian time-domain likelihood is not computationally feasible for large $n$ for the ARFIMA and ARTFIMA models. Further, no such comparison can be made in the ARTFIMA-SV case as the model is non-Gaussian \cite{BREIDT1998325}.

\begin{figure}[H]
        \centering
        \includegraphics[width=0.50\textwidth]{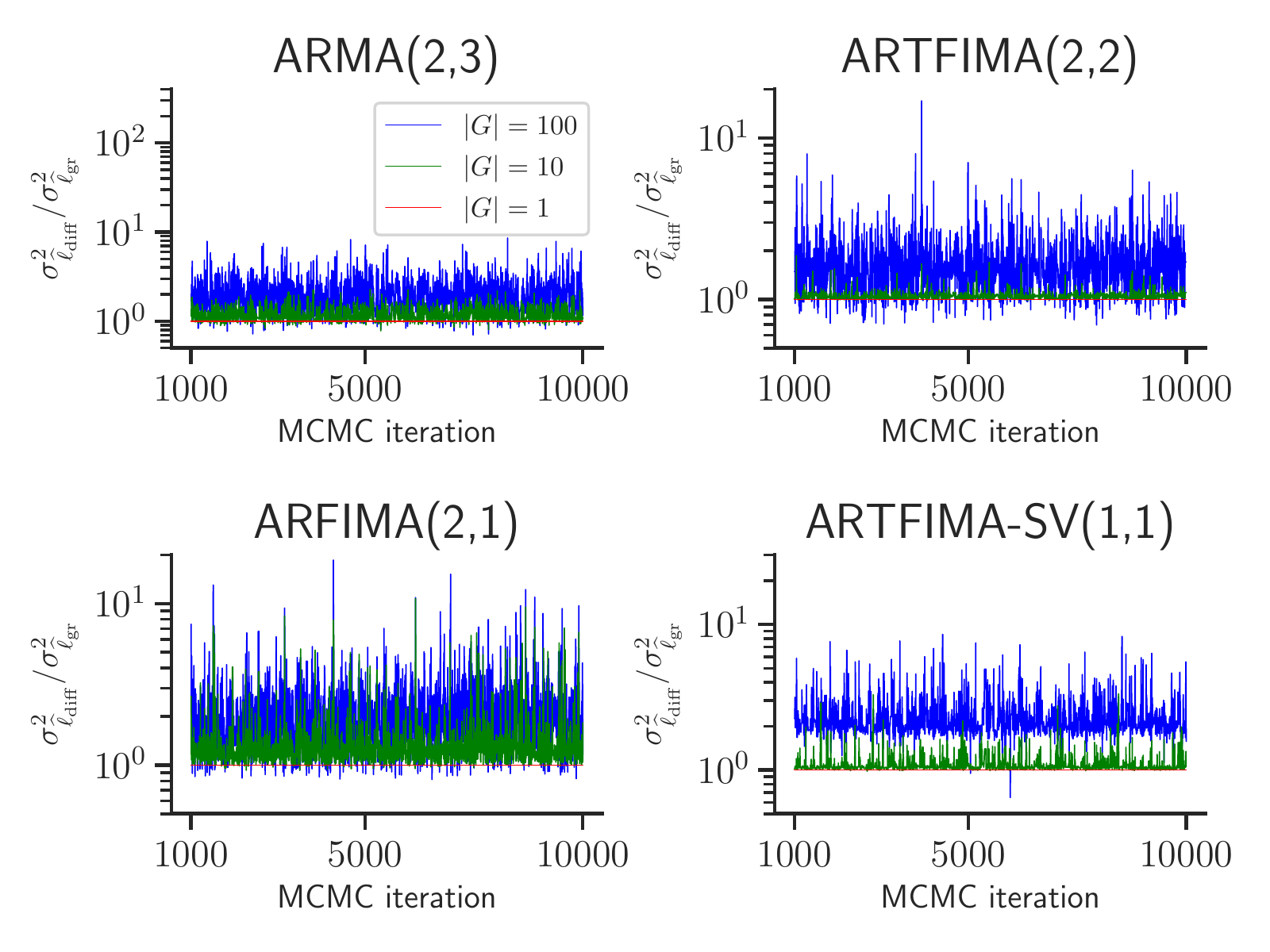}
        \caption{Effect of grouping the Taylor series control variate for all examples. The figure shows the relative variance reduction (in log-scale) with no grouping (i.e. one observation per group $|G|=1$) as baseline over the MCMC iterations post burn-in. A value larger than 1 means that the corresponding control variate is more efficient compared to the baseline.} \label{fig:EffectOfGrouping}
    \end{figure}

         Figure \ref{fig:EffectOfGrouping} shows that in general the use of grouping in the control variates reduces the variance, especially for the more complex models. This experiment is performed using the last $2001$ time observations for each model ($n=1000$ in the frequency domain) to prevent $||\v \theta - \v \theta^\star||$ becoming too small.

\begin{figure}[ht]
        \centering
        \includegraphics[width=0.50\textwidth]{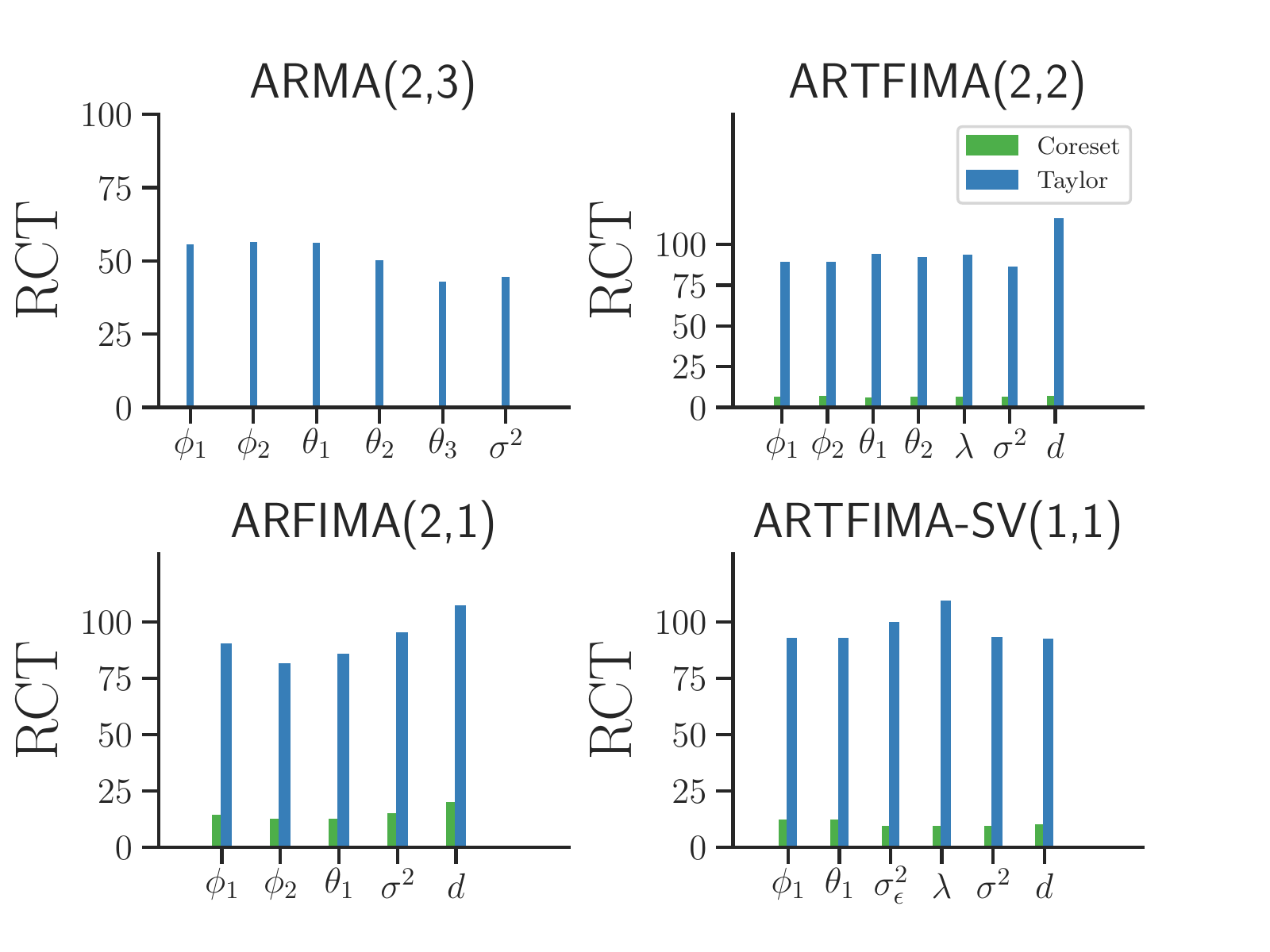}
        \caption{Relative computational time each parameter in all examples. Results are relative to full-data MCMC (larger is better for Subsampling MCMC). }
        \label{fig:ARMArct}
    \end{figure}

   Figure \ref{fig:ARMArct} reports the relative computational time with MCMC on the full-data Whittle likelihood as baseline for all parameters --- showing that our method introduces close to two orders of magnitude speedup on these examples.

\begin{figure}[ht]
        \centering
                \includegraphics[width=0.50\textwidth]{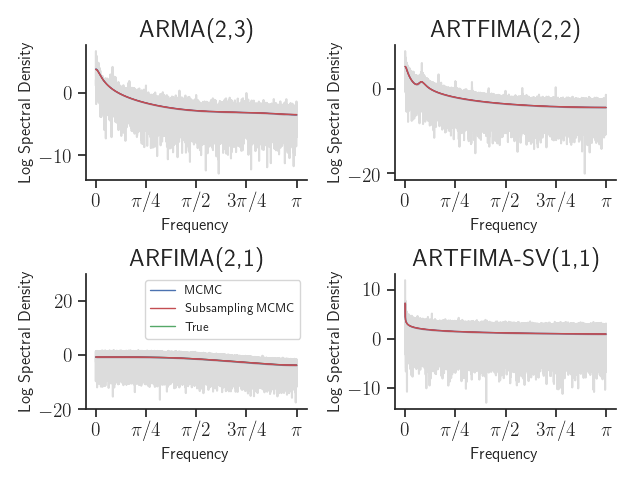}
        \caption{Periodogram data (grey) in log-scale with posterior means of the (log) spectral density estimate from MCMC on full-data Whittle likelihood and from Subsampling MCMC on the subsampled Whittle likelihood for all four models. The figure also shows the true spectral density, which is only available for the ARFIMA(2,1) example (simulated data). All posterior means are indistinguishable, showing the accuracy of our method and, moreover, the accuracy of the Whittle approximation for the ARFIMA(2,1) example.} \label{fig:SpectralDensity}
    \end{figure}

         Finally, Figure \ref{fig:SpectralDensity} plots the periodogram and posterior mean spectral density for all models using our Subsampling MCMC method and, for comparison, the full-data MCMC (on the Whittle likelihood) method. The figure confirms the accuracy of our method and, moreover, that we recover the true spectral density perfectly in a simulated data setting (ARFIMA(2,1)). Recall that due to \eqref{eq:theWhittleLikelihood}, the fitting of a Stationary time series model is essentially a univariate regression problem for the spectral density. Thus, the figure also shows that the models we consider capture features of the real world datasets while avoiding overfitting.

    \section{Discussion}\label{sec:Discussion}
    We introduce novel methods allowing efficient Bayesian inference in stationary models for large time series datasets. The idea is simple and elegant: overcome the lack of independence in the data by transforming them to the frequency domain where they are independent. This work is, to our knowledge, the first to extend scalable MCMC algorithms to models with temporal dependence. The article is focused on Subsampling MCMC, but the ideas introduced here can be directly applied to many other scalable inference approaches, e.g.,those in  \citet{Quiroz2016, quiroz2018delayed} and \citet{Cornish}. Subsampling of periodogram frequencies also extends beyond the MCMC setting, for example to {\em Doubly} Stochastic Variational Inference \citep{titsias2014doubly}. We also introduce novel control variate schemes based on grouping and coresets to improve the  robustness of Subsampling MCMC. 
    
    Two immediate and interesting extensions of our methods are to multiple time series and locally stationary time series; \citet{dahlhaus2000likelihood} define a suitable analogue to the univariate Whittle likelihood for these cases. A third extension is to semi-parametric and non-parametric spectral density estimation as in for example \citet{Carter1997}, \citet{Choudhuri2004}, and \citet{Edwards2019}. Finally, it has not escaped our notice that our approach can be directly extended to spatial and spatio-temporal data using the multidimensional DFT (see \citet{SpatialAsymptotics} for results regarding the asymptotic distribution of the Fourier transform in the spatial setting).

\section*{Acknowledgements}
Salomone and Kohn were partially supported by the Australian Centre of Excellence for Mathematical and Statistical Frontiers, under Australian Research Council grant CE140100049. Villani was supported by the Swedish Foundation for Strategic Research (Smart Systems: RIT 15-0097).
\bibliography{whittleRef}

\begin{thebibliography}{39}
\providecommand{\natexlab}[1]{#1}
\providecommand{\url}[1]{\texttt{#1}}
\expandafter\ifx\csname urlstyle\endcsname\relax
  \providecommand{\doi}[1]{doi: #1}\else
  \providecommand{\doi}{doi: \begingroup \urlstyle{rm}\Url}\fi

\bibitem[Andrieu et~al.(2009)Andrieu, Roberts, et~al.]{Andrieu2009}
Andrieu, C., Roberts, G.~O., et~al.
\newblock The pseudo-marginal approach for efficient {M}onte {C}arlo
  computations.
\newblock \emph{The Annals of Statistics}, 37\penalty0 (2):\penalty0 697--725,
  2009.

\bibitem[Bardenet et~al.(2017)Bardenet, Doucet, and Holmes]{bardenet2017markov}
Bardenet, R., Doucet, A., and Holmes, C.
\newblock On {M}arkov chain {M}onte {C}arlo methods for tall data.
\newblock \emph{The Journal of Machine Learning Research}, 18\penalty0
  (1):\penalty0 1515--1557, 2017.

\bibitem[Barndorff-Nielsen \& Schou(1973)Barndorff-Nielsen and
  Schou]{Barndorff-Nielsen1973}
Barndorff-Nielsen, O. and Schou, G.
\newblock On the parametrization of autoregressive models by partial
  autocorrelations.
\newblock \emph{Journal of Multivariate Analysis}, 3\penalty0 (4):\penalty0
  408--419, 1973.

\bibitem[Blei et~al.(2017)Blei, Kucukelbir, and McAuliffe]{blei2017variational}
Blei, D.~M., Kucukelbir, A., and McAuliffe, J.~D.
\newblock Variational inference: A review for statisticians.
\newblock \emph{Journal of the American Statistical Association}, 112\penalty0
  (518):\penalty0 859--877, 2017.

\bibitem[Breidt et~al.(1998)Breidt, Crato, and de~Lima]{BREIDT1998325}
Breidt, F., Crato, N., and de~Lima, P.
\newblock The detection and estimation of long memory in stochastic volatility.
\newblock \emph{Journal of Econometrics}, 83\penalty0 (1):\penalty0 325 -- 348,
  1998.

\bibitem[Campbell \& Beronov(2019)Campbell and Beronov]{Campbell2019}
Campbell, T. and Beronov, B.
\newblock Sparse variational inference: {B}ayesian coresets from scratch.
\newblock \emph{arXiv preprint arXiv:1906.03329}, 2019.

\bibitem[Campbell \& Broderick(2018)Campbell and Broderick]{Campbell2018}
Campbell, T. and Broderick, T.
\newblock {B}ayesian coreset construction via greedy iterative geodesic ascent.
\newblock In \emph{Proceedings of the 35th International Conference on Machine
  Learning}, pp.\  698--706, 2018.

\bibitem[Campbell \& Broderick(2019)Campbell and Broderick]{Campbell2019a}
Campbell, T. and Broderick, T.
\newblock Automated scalable {B}ayesian inference via {H}ilbert coresets.
\newblock \emph{The Journal of Machine Learning Research}, 20\penalty0
  (1):\penalty0 551--588, 2019.

\bibitem[Carter \& Kohn(1997)Carter and Kohn]{Carter1997}
Carter, C.~K. and Kohn, R.
\newblock Semiparametric {B}ayesian inference for time series with mixed
  spectra.
\newblock \emph{Journal of the Royal Statistical Society: Series B (Statistical
  Methodology)}, 59\penalty0 (1):\penalty0 255--268, 1997.

\bibitem[Chan \& Palma(1998)Chan and Palma]{chan1998state}
Chan, N.~H. and Palma, W.
\newblock State space modeling of long-memory processes.
\newblock \emph{Annals of Statistics}, pp.\  719--740, 1998.

\bibitem[Chen et~al.(2014)Chen, Fox, and Guestrin]{SG-HMC}
Chen, T., Fox, E., and Guestrin, C.
\newblock Stochastic gradient {H}amiltonian monte carlo.
\newblock In \emph{International conference on machine learning}, pp.\
  1683--1691, 2014.

\bibitem[Choudhuri et~al.(2004)Choudhuri, Ghosal, and Roy]{Choudhuri2004}
Choudhuri, N., Ghosal, S., and Roy, A.
\newblock Bayesian estimation of the spectral density of a time series.
\newblock \emph{Journal of the American Statistical Association}, 99\penalty0
  (468):\penalty0 1050--1059, 2004.

\bibitem[Cornish et~al.(2019)Cornish, Vanetti, Bouchard{-}C{\^{o}}t{\'{e}},
  Deligiannidis, and Doucet]{Cornish}
Cornish, R., Vanetti, P., Bouchard{-}C{\^{o}}t{\'{e}}, A., Deligiannidis, G.,
  and Doucet, A.
\newblock Scalable {M}etropolis-{H}astings for exact {B}ayesian inference with
  large datasets.
\newblock In \emph{Proceedings of the 36th International Conference on Machine
  Learning, {ICML} 2019}, pp.\  1351--1360, 2019.

\bibitem[Dahlhaus et~al.(2000)]{dahlhaus2000likelihood}
Dahlhaus, R. et~al.
\newblock A likelihood approximation for locally stationary processes.
\newblock \emph{The Annals of Statistics}, 28\penalty0 (6):\penalty0
  1762--1794, 2000.

\bibitem[Dang et~al.(2019)Dang, Quiroz, Kohn, Tran, and
  Villani]{dang2019hamiltonian}
Dang, K.-D., Quiroz, M., Kohn, R., Tran, M.-N., and Villani, M.
\newblock Hamiltonian {M}onte {C}arlo with energy conserving subsampling.
\newblock \emph{Journal of Machine Learning Research}, 20\penalty0
  (100):\penalty0 1--31, 2019.

\bibitem[Deligiannidis et~al.(2018)Deligiannidis, Doucet, and
  Pitt]{deligiannidis2018correlated}
Deligiannidis, G., Doucet, A., and Pitt, M.~K.
\newblock The correlated pseudomarginal method.
\newblock \emph{Journal of the Royal Statistical Society: Series B (Statistical
  Methodology)}, 80\penalty0 (5):\penalty0 839--870, 2018.

\bibitem[Edwards et~al.(2019)Edwards, Meyer, and Christensen]{Edwards2019}
Edwards, M.~C., Meyer, R., and Christensen, N.
\newblock Bayesian nonparametric spectral density estimation using {B}-spline
  priors.
\newblock \emph{Statistics and Computing}, 29\penalty0 (1):\penalty0 67--78,
  Jan 2019.

\bibitem[Fox \& Taqqu(1986)Fox and Taqqu]{FoxTaqqu1986}
Fox, R. and Taqqu, M.
\newblock Large-sample properties of parameter estimates for strongly dependent
  stationary {Gaussian} time series.
\newblock \emph{The Annals of Statistics}, 14\penalty0 (2):\penalty0 517--532,
  1986.

\bibitem[Granger \& Joyeux(1980)Granger and Joyeux]{Granger1980}
Granger, C.~W. and Joyeux, R.
\newblock An introduction to long-memory time series models and fractional
  differencing.
\newblock \emph{Journal of time series analysis}, 1\penalty0 (1):\penalty0
  15--29, 1980.

\bibitem[Huggins et~al.(2016)Huggins, Campbell, and Broderick]{Huggins2016}
Huggins, J., Campbell, T., and Broderick, T.
\newblock Coresets for scalable {B}ayesian logistic regression.
\newblock In \emph{Advances in Neural Information Processing Systems}, pp.\
  4080--4088, 2016.

\bibitem[Hyndman \& Khandakar(2008)Hyndman and Khandakar]{ForecastPackage}
Hyndman, R. and Khandakar, Y.
\newblock Automatic time series forecasting: The forecast package for {R}.
\newblock \emph{Journal of Statistical Software}, 27\penalty0 (3):\penalty0
  1--22, 2008.

\bibitem[Lindgren(2012)]{lindgren2012stationary}
Lindgren, G.
\newblock \emph{Stationary stochastic processes: theory and applications}.
\newblock Chapman and Hall/CRC, 2012.

\bibitem[Marin et~al.(2012)Marin, Pudlo, Robert, and Ryder]{Marin2012}
Marin, J.-M., Pudlo, P., Robert, C.~P., and Ryder, R.~J.
\newblock Approximate {B}ayesian computational methods.
\newblock \emph{Statistics and Computing}, 22\penalty0 (6):\penalty0
  1167--1180, Nov 2012.

\bibitem[Peligrad \& Zhang(2017)Peligrad and Zhang]{SpatialAsymptotics}
Peligrad, M. and Zhang, N.
\newblock Central limit theorem for {F}ourier transform and periodogram of
  random fields.
\newblock \emph{Bernoulli}, 25, 2017.

\bibitem[Peligrad et~al.(2010)Peligrad, Wu, et~al.]{peligrad2010central}
Peligrad, M., Wu, W.~B., et~al.
\newblock Central limit theorem for {F}ourier transforms of stationary
  processes.
\newblock \emph{The Annals of Probability}, 38\penalty0 (5):\penalty0
  2009--2022, 2010.

\bibitem[Plummer et~al.(2006)Plummer, Best, Cowles, and Vines]{plummer2006coda}
Plummer, M., Best, N., Cowles, K., and Vines, K.
\newblock {CODA}: {C}onvergence diagnosis and output analysis for {MCMC}.
\newblock \emph{R News}, 6\penalty0 (1):\penalty0 7--11, 2006.

\bibitem[Quiroz et~al.(2018{\natexlab{a}})Quiroz, Tran, Villani, and
  Kohn]{quiroz2018delayed}
Quiroz, M., Tran, M.-N., Villani, M., and Kohn, R.
\newblock Speeding up {MCMC} by delayed acceptance and data subsampling.
\newblock \emph{Journal of Computational and Graphical Statistics}, 27\penalty0
  (1):\penalty0 12--22, 2018{\natexlab{a}}.

\bibitem[Quiroz et~al.(2018{\natexlab{b}})Quiroz, Tran, Villani, Kohn, and
  Dang]{Quiroz2016}
Quiroz, M., Tran, M.-N., Villani, M., Kohn, R., and Dang, K.-D.
\newblock The block-{P}oisson estimator for optimally tuned exact subsampling
  {MCMC}.
\newblock \emph{arXiv preprint arXiv:1603.08232v5}, 2018{\natexlab{b}}.

\bibitem[Quiroz et~al.(2019{\natexlab{a}})Quiroz, Kohn, Villani, and
  Tran]{quiroz2019speeding}
Quiroz, M., Kohn, R., Villani, M., and Tran, M.-N.
\newblock Speeding up {MCMC} by efficient data subsampling.
\newblock \emph{Journal of the American Statistical Association}, 114\penalty0
  (526):\penalty0 831--843, 2019{\natexlab{a}}.

\bibitem[Quiroz et~al.(2019{\natexlab{b}})Quiroz, Villani, Kohn, Tran, and
  Dang]{quiroz2018subsampling}
Quiroz, M., Villani, M., Kohn, R., Tran, M.-N., and Dang, K.-D.
\newblock Subsampling {MCMC} - {A}n introduction for the survey statistician.
\newblock \emph{Sankhya A}, 80\penalty0 (1):\penalty0 33--69,
  2019{\natexlab{b}}.

\bibitem[Rue et~al.(2009)Rue, Martino, and Chopin]{rue2009approximate}
Rue, H., Martino, S., and Chopin, N.
\newblock Approximate {B}ayesian inference for latent {G}aussian models by
  using integrated nested {L}aplace approximations.
\newblock \emph{Journal of the Royal Statistical Society: Series B (Statistical
  Methodology)}, 71\penalty0 (2):\penalty0 319--392, 2009.

\bibitem[Sabzikar et~al.(2019)Sabzikar, McLeod, and Meerschaert]{Sabzikar2019}
Sabzikar, F., McLeod, A.~I., and Meerschaert, M.~M.
\newblock Parameter estimation for {ARTFIMA} time series.
\newblock \emph{Journal of Statistical Planning and Inference}, 200:\penalty0
  129 -- 145, 2019.

\bibitem[Shao et~al.(2007)Shao, Wu, et~al.]{Shao2007}
Shao, X., Wu, W.~B., et~al.
\newblock Asymptotic spectral theory for nonlinear time series.
\newblock \emph{The Annals of Statistics}, 35\penalty0 (4):\penalty0
  1773--1801, 2007.

\bibitem[Sykulski et~al.(2019)Sykulski, Guillaumin, Olhede, Early, and
  Lilly]{Sykulski2019}
Sykulski, A.~M., Guillaumin, A.~P., Olhede, S.~C., Early, J.~J., and Lilly,
  J.~M.
\newblock {The debiased Whittle likelihood}.
\newblock \emph{Biometrika}, 106\penalty0 (2):\penalty0 251--266, 2019.

\bibitem[Tamaki(2008)]{Tamaki2008}
Tamaki, K.
\newblock The {B}ernstein-von {M}ises theorem for stationary processes.
\newblock \emph{Journal of the Japan Statistical Society}, 38\penalty0
  (2):\penalty0 311--323, 2008.

\bibitem[Titsias \& L{\'a}zaro-Gredilla(2014)Titsias and
  L{\'a}zaro-Gredilla]{titsias2014doubly}
Titsias, M. and L{\'a}zaro-Gredilla, M.
\newblock Doubly stochastic variational {B}ayes for non-conjugate inference.
\newblock In \emph{International Conference on Machine Learning}, pp.\
  1971--1979, 2014.

\bibitem[Tran et~al.(2016)Tran, Kohn, Quiroz, and Villani]{Tran2016block}
Tran, M.-N., Kohn, R., Quiroz, M., and Villani, M.
\newblock The block pseudo-marginal sampler.
\newblock \emph{arXiv preprint arXiv:1603.02485}, 2016.

\bibitem[Welling \& Teh(2011)Welling and Teh]{SGLD}
Welling, M. and Teh, Y.~W.
\newblock Bayesian learning via stochastic gradient {L}angevin dynamics.
\newblock In \emph{Proceedings of the 28th international conference on machine
  learning (ICML-11)}, pp.\  681--688, 2011.

\bibitem[Whittle(1953)]{Whittle1953}
Whittle, P.
\newblock Estimation and information in stationary time series.
\newblock \emph{Arkiv f{\"o}r matematik}, 2\penalty0 (5):\penalty0 423--434,
  1953.

\end{thebibliography}
\bibliographystyle{icml2020}

\clearpage
\appendix 
\section{Supplementary Material}
 \subsection{Asymptotic Analysis\label{subsec:Asymptotic}}
     Let $\v \theta^\star$ be a mode of $\pi_{W}(\v \theta)$ and $\Delta(\v \theta):=\partial^2\log\pi_W(\v \theta)/\partial\v \theta\partial\v \theta^\top$. As in \citet{quiroz2018subsampling}, we need the following regularity conditions on the Whittle likelihood, which can be justified by the asymptotic normality of the maximum Whittle-likelihood estimator \cite{FoxTaqqu1986} and the Bernstein-von Mises theorem for the Whittle measure \cite{Tamaki2008}. 
    
    \begin{assumption}\label{ass1}
    \begin{itemize}
    \item[(A1)] For each $i$, $$\ell_{W,i}(\v \theta):=-\log f_{\v \theta}(\omega_i)- {\cal I}(\omega_i)/f_{\v \theta}(\omega_i)$$ is three times differentiable with
    $\max_{j,k,l \in \{1, \dots, p\}} \sup_{\v \theta \in \v \theta} \Biggr \rvert \frac{\partial^3 \ell_{W,i}(\v \theta)}
    {\partial \v \theta_j\partial \v \theta_k\partial \v \theta_l}\Biggr \rvert
    $ bounded.
    \item[(A2)] $\Delta(\v \theta^\star)$ is negative definite.
    \item[(A3)] $\|\Sigma\|_2=O(n^{-1})$, where  $\Sigma=\big(-\Delta(\v \theta^\star)\big)^{-1}$.
    \item[(A4)] For any $\epsilon>0$, there exist a $\delta_\epsilon>0$ and an integer $N_{1,\epsilon}$ such that for any $n>N_{1,\epsilon}$ and $\v \theta\in H(\v \theta^\star,\delta_\epsilon)$, $\Delta(\v \theta)$ exists and satisfies
    \[-A(\epsilon)\leq \Delta(\v \theta)\big(\Delta(\v \theta^\star)\big)^{-1}-I\leq A(\epsilon)\]
    where $A(\epsilon)$ is a positive semidefinite matrix whose largest eigenvalue goes to 0 as $\epsilon\to0$.
    \item[(A5)]  For any $\delta>0$, there exists a positive integer $N_{2,\delta}$ and two positive numbers $c$ and $\kappa$ such that
    for $n>N_{2,\delta}$ and $\v \theta\not\in H(\v \theta^\star,\delta)$
    \[\frac{\pi_{W}(\v \theta)}{\pi_{W}(\v \theta^\star)}<\exp\left(-c\big[(\v \theta-\v \theta^\star)^\top\Sigma^{-1}(\v \theta-\v \theta^\star)\big]^\kappa\right).\]
    \end{itemize}
    \end{assumption}

\section{Additional results}
\begin{figure}[ht]
        \centering
        \includegraphics[width=0.50\textwidth]{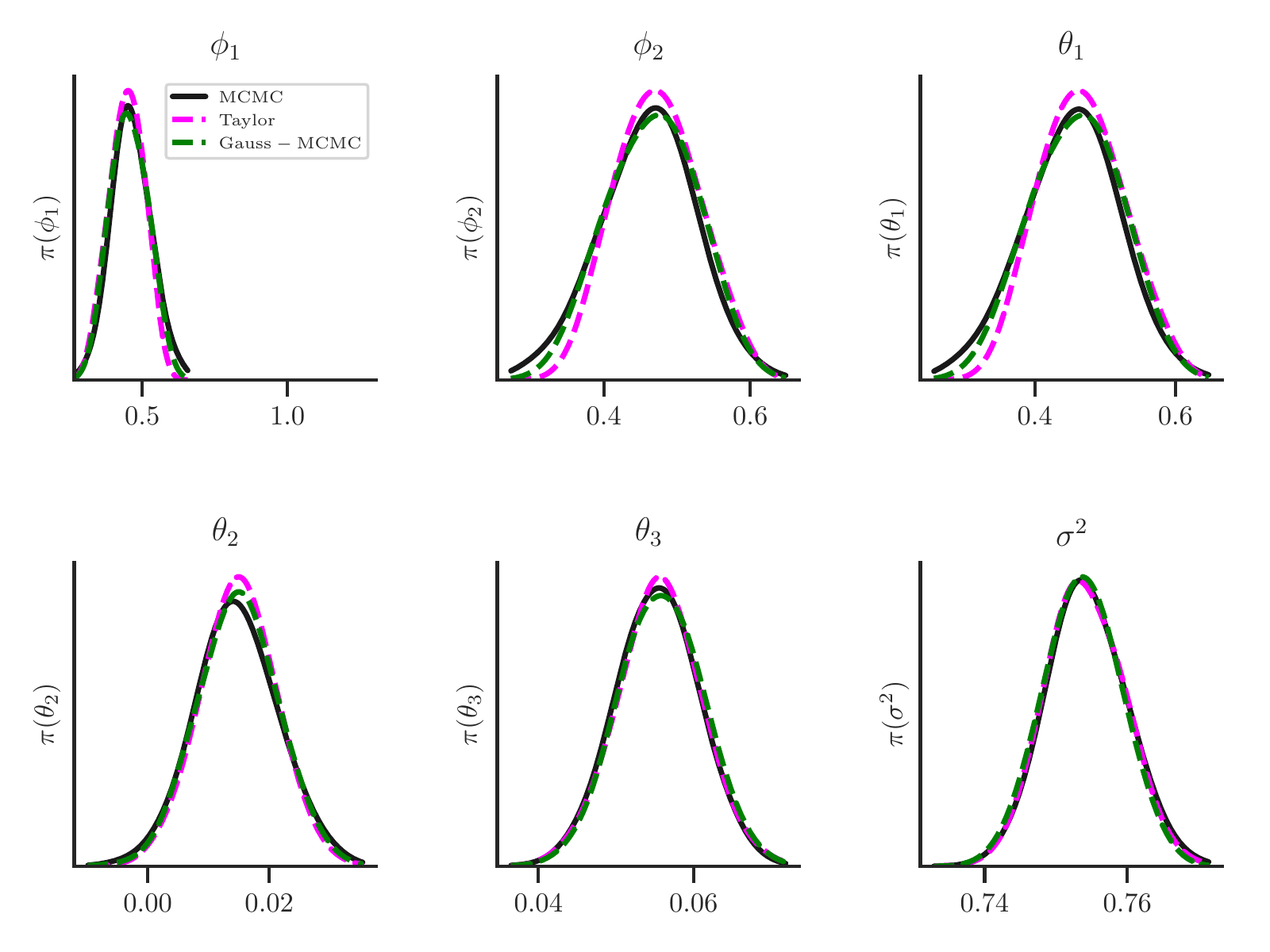}
        \caption{Kernel Density Estimates of all marginal distributions for the ARMA(2,3) example. MCMC is the full-data MCMC on the Whittle likelihood. Taylor is the Subsampling MCMC method using the Taylor series expanded control variates. Gauss-MCMC is the full-data MCMC on the time domain Gaussian likelihood. } \label{fig:kde_all_params_ARMA}
    \end{figure}

\begin{figure}[ht]
        \centering
        \includegraphics[width=0.50\textwidth]{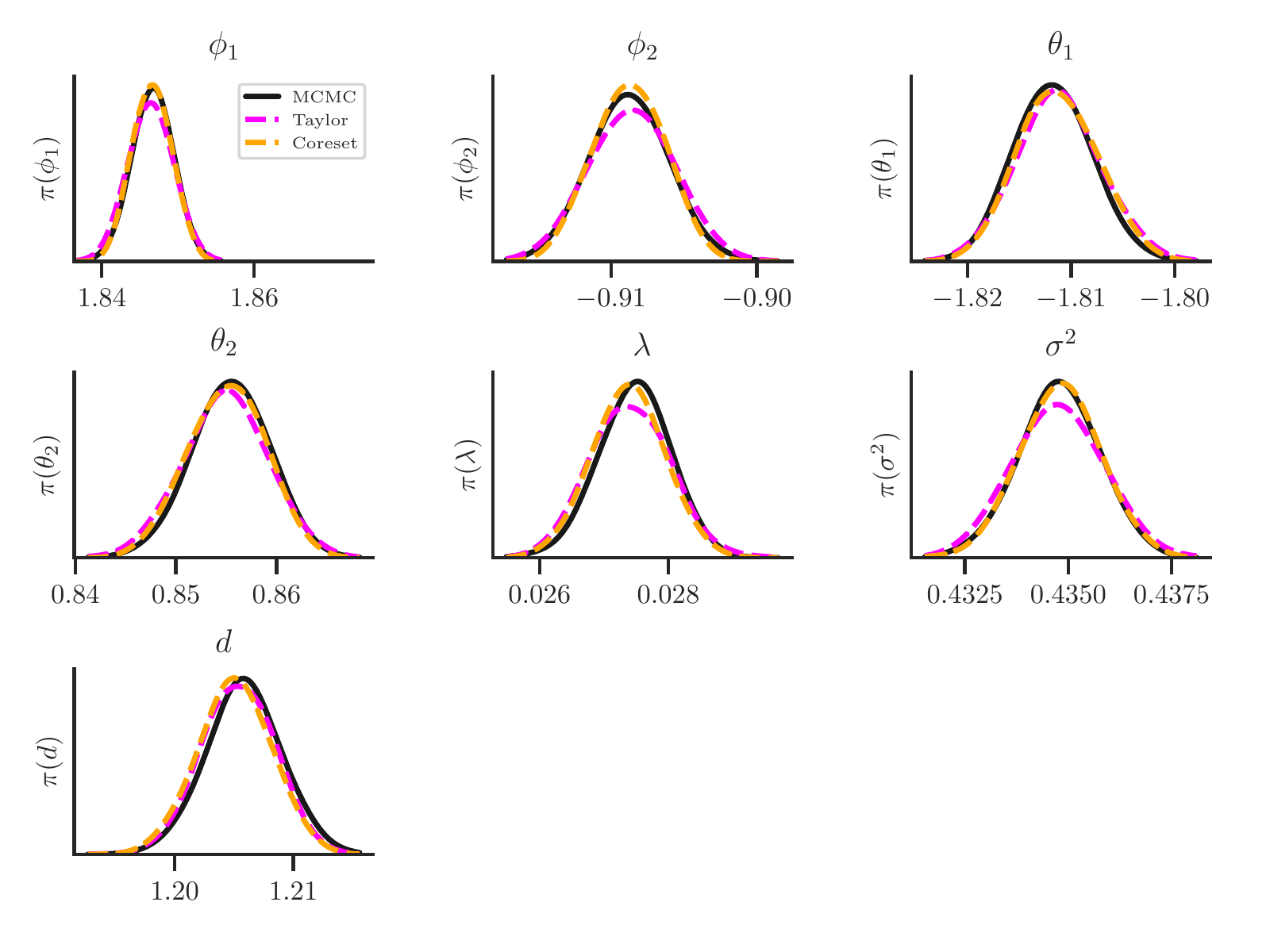}
        \caption{Kernel Density Estimates of all marginal distributions for the ARTFIMA(2,2) example.} \label{fig:kde_all_params_ARTFIMA}
    \end{figure}

\begin{figure}[ht]
        \centering
        \includegraphics[width=0.50\textwidth]{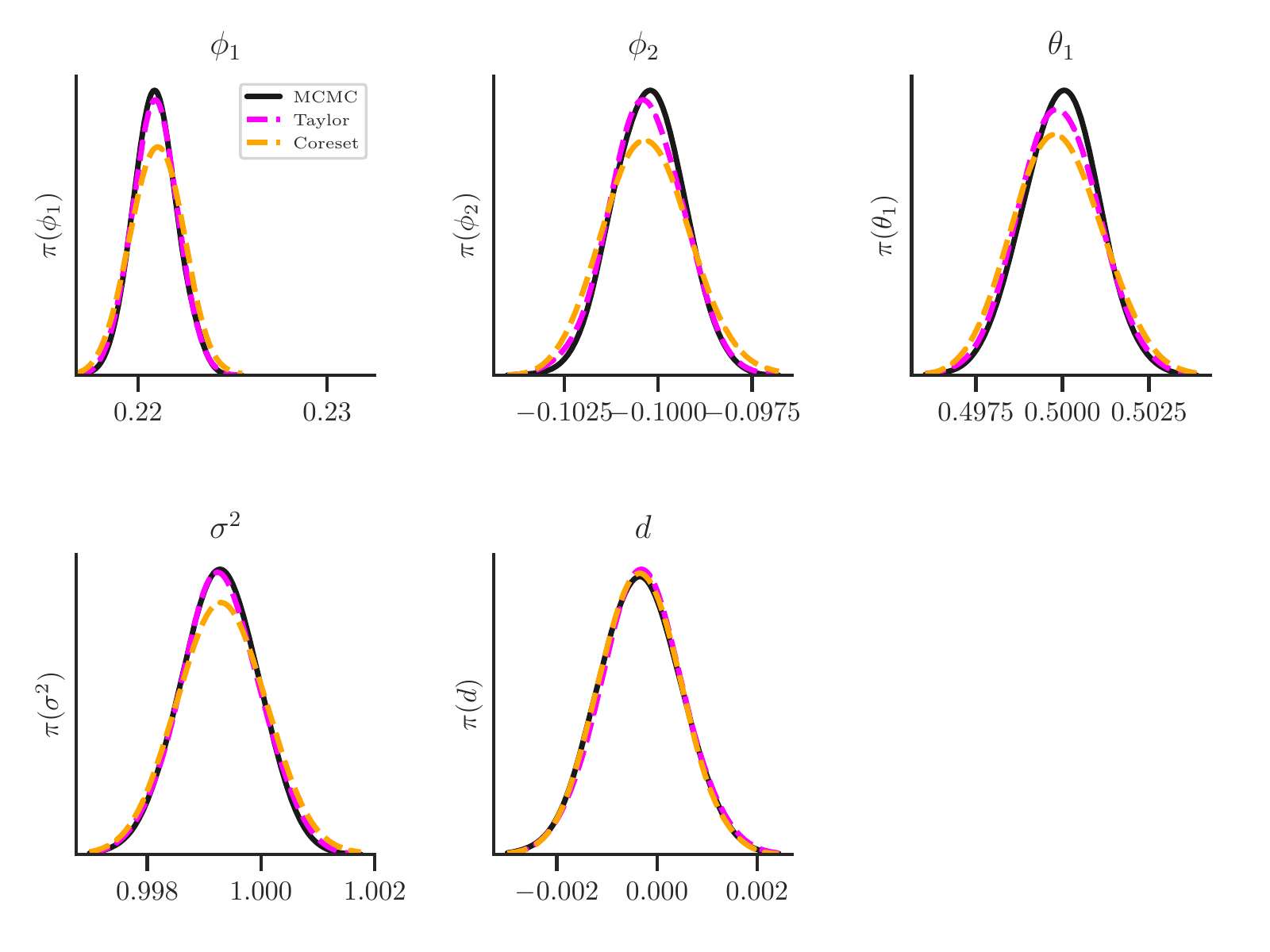}
        \caption{Kernel Density Estimates of all marginal distributions for the ARFIMA(2,1) example. MCMC is the full-data MCMC on the Whittle likelihood. Taylor and coreset are the Subsampling MCMC methods using the corresponding control variates.} \label{fig:kde_all_params_ARFIMA}
    \end{figure}

\begin{figure}[ht]
        \centering
        \includegraphics[width=0.50\textwidth]{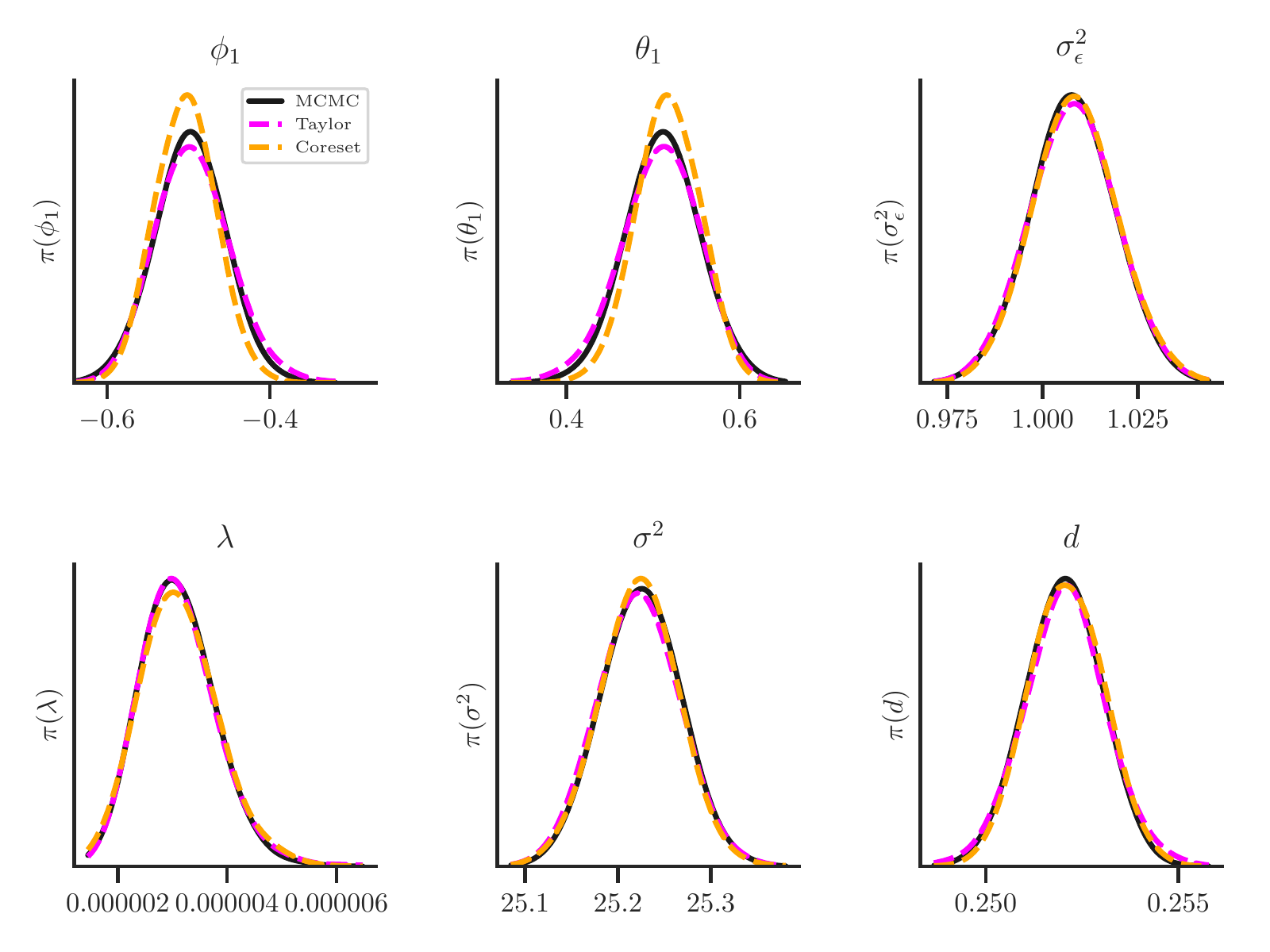}
        \caption{Kernel Density Estimates of all marginal distributions for the ARTFIMA-SV(1,1) example. MCMC is the full-data MCMC on the Whittle likelihood. Taylor and coreset are the Subsampling MCMC methods using the corresponding control variates.} \label{fig:kde_all_params_ARTFIMA-SV}
    \end{figure}

The accuracy of all marginal posteriors is illustrated for each example in Figures \ref{fig:kde_all_params_ARMA}-\ref{fig:kde_all_params_ARTFIMA-SV}; see the captions for details. The Taylor control variates is very accurate or close to very accurate for all parameters. The coreset control variate is often accurate, however, for some of the parameters the bias is noticeable. This is because the variance of the coreset control variate in the examples considered is larger than that of the Taylor series control variate. We expect the coreset control variate to outperform the Taylor series control variate in examples where the grouped data density is multimodal on the logarithmic scale. This case is clearly outside the scope of the Taylor series control variate, since it assumes that the grouped data density is approximately quadratic on the logarithmic scale. We leave this for future research. 

\end{document}